\documentclass[conference,10pt]{IEEEtran}

\usepackage[latin1]{inputenc}
\usepackage{amsmath}
\usepackage{amssymb}
\usepackage{epsfig}
\usepackage{url}
\usepackage[ruled,lined,linesnumbered]{algorithm2e}
\usepackage{eso-pic}
\newtheorem{fact}{Fact}
\newtheorem{theorem}{Theorem}
\DeclareMathOperator*{\argmax}{argmax}
\def\c{{\cal C}}
\def\d{{\cal D}}
\def\x{{\cal X}}
\newcommand{\remove}[1]{}

\AddToShipoutPicture*{\small \sffamily\raisebox{1.8cm}{\hspace{1.8cm}978-1-4244-8998-5/10/\$26.00 \copyright\ 2010 IEEE}}

\begin{document}
%
\title{Community Detection via\\ Semi--Synchronous Label Propagation Algorithms}

\author{\IEEEauthorblockN{Gennaro Cordasco and Luisa Gargano}
\IEEEauthorblockA{Dipartimento di Informatica  ed Applicazioni ``R.M. Capocelli''\\ 
University of Salerno,\\
Fisciano 84084, ITALY \\
\texttt{\{cordasco,lg\}@dia.unisa.it}} 
}

\maketitle

\begin{abstract}
A recently introduced  novel community detection strategy is based on a label propagation algorithm (LPA) which uses the diffusion of information in the network to identify communities. Studies of LPAs showed that the strategy is effective in finding a good community structure.
Label propagation step can be performed in parallel on all nodes (synchronous model) or sequentially (asynchronous model);  both  models present some drawback, e.g., algorithm termination is nor granted in the first case, performances can be  worst in the second case.
In this paper, we present a semi--synchronous version of LPA which aims to combine the advantages of both synchronous and asynchronous models. We prove that our models always converge to a stable labeling. 
Moreover, we experimentally investigate the effectiveness of the proposed 
strategy comparing its performance with the asynchronous model both in terms of quality, efficiency and stability.
Tests show that the proposed protocol does not harm the quality of the partitioning. Moreover it is quite efficient; each propagation step is extremely parallelizable and it is more stable than the asynchronous model, thanks to the fact that only a small amount of randomization is used by our proposal.
\end{abstract}

\section{Introduction}
Collaboration networks, the Internet, the World-Wide-Web, Biological networks, Communication and Transport networks, Social networks are just some examples of wide complex networks.  An interesting property to investigate, typical to many  such networks, is the community structure, i.e. the division of networks into groups  of nodes that are similar among them but dissimilar from the rest of the network. The capability of detecting the partitioning of a network into communities can give important insights into the organization and behavior of the system that the network models.

The generally adopted notion of community structure in complex networks \cite{GN02} refers to the fact that nodes in many real networks appear to group into subgraphs which are densely inter-connected and sparsely connected to other parts of the network. 
The quite challenging problem of community detection has attracted ample attention in recent years and
community detection methods have been applied in a wide range of scientific problems such as Social networks,  Citation networks,  the World Wide Web, and many others \cite{AB02}.
Several  different approaches have been proposed to find community structures in networks;  reviews of the various methods present in the literature can be found in 
\cite{DDDA05} and \cite{OL09}.

Recently Raghavan  et al.  \cite{RAK07} proposed   a label propagation algorithm (LPA)  for detecting network communities. 
This algorithm uses only the network structure as a guide, and can be summarized as follows: Each node in the network is first given a unique label; at each iteration, each node is updated by choosing the label which is the most frequent among its neighbors -- if  multiple choices are possible (as for example  in the beginning), one label is picked randomly. 
Experiments  have shown that the label propagation technique is very effective  in discovering accurate community structure. \\
In \cite{RAK07},  the authors suggest to use an asynchronous label propagation approach, since otherwise the process may result in a cyclic oscillation of the labels of some vertices which precludes the convergence (termination) of the algorithm;  the use of an asynchronous algorithm means to update only 
one node at time which implies $O(m)$ time for each iteration, if the network has $m$ links.
In general, while the experiments reported in \cite{LHLC09} empirically show that the expected number of iterations grows logarithmically with respect to the size of the network, the problem of  analytically proving the convergence of this method and of determining its speed is still an open problem  (in both synchronous and asynchronous models). 

Liu and Murata \cite{LM09} introduced a variation of the label propagation algorithm specifically developed for bipartite network; the algorithm is semi--synchronous: it allows to update simultaneously all the nodes belonging to one of the partitions of the bipartite networks.  The authors experience that by running this version of the LPA, the oscillation problem disappears. However, no formal proof (or an informal hint) that shows why their proposal is able to defeat the oscillation issue has been provided.

In this paper we extend the approach in \cite{LM09}  to any graph. Moreover, we show that no randomization is needed for obtaining good algorithm performance and formally prove that the proposed algorithm 
converges.

The rest of the paper is organized as follows. Section \ref{not} describes the notation used in the paper and introduce the modularity measure that will be used to evaluate the quality of the discovered community structures. In Section \ref{sec:RW}, we briefly discuss on community detection strategies and introduce the LPA. Section \ref{stopCriteria} introduces three variants of the standard LPA, which have been proposed to improve its features and discusses the stop criteria of LPAs.
In Section \ref{sec:ourResult}, we present and analyze our algorithm. In Section \ref{sec:Experiments}, we report on the experimental data.

\section{Notation: Graphs, Partitioning and Coloring} \label{not}
Let $G=(V,E)$ be an undirected connected network having $n=|V|$ vertices and $m=|E|$ edges. Let $v\in V$, we denote by  $$N(v)=\{u  \ \colon \ u\in V,\ \{u,v\}\in E\}$$ the neighborhood of $v$, by $$deg(v)=\lvert N(v) \rvert$$ the degree of $v$ and by $$deg(G)=\max_{v\in V} deg(v)$$ the maximum degree over all the vertices in $G$.

\bigskip
Denote by $\c=\{C_1,C_2,\ldots,C_k\}$ a partition of the vertices in $V$. Each $C_i$ is called community. Each community $C_i$ is associated to an induced subgraph $G(C_i)=(C_i, E(C_i))$ of $G$, where  $E(C_i)=\{\{u,v\}  \ \colon \ u,v \in C_i, \, \{u,v\} \in E \}.$ 

A good community detection strategy should strive to achieve two conflicting goals: \begin{itemize}
\item provide an accurate network partition and 
\item keep low the computational complexity of the algorithm so as the algorithm can be applied to very large networks. 
\end{itemize}
While the latter goal is easily measured by means of a standard worldwide recognized approach (e.g.,  Asymptotic notation),  how to evaluate the quality of a given network partition is a recently raised question. 
Several metrics have been proposed in the literature, e.g.,  edges density  or NMI (Normalized mutual information) \cite{DDDA05}.
Among these, a metric which has been widely adopted to measure the accuracy of a network partition is the graph modularity  proposed by Newman \cite{NG04}. This measure is obtained by summing up, over all communities of a given partition,  the difference between the observed fraction of links inside the community and the expected value for a null model, that is, a randomized network having the same size and same degree sequence. Formally, the modularity of a given partition $\c$ of a graph $G=(V,E)$ is defined as 
\begin{equation}
q(\c)= \sum_{C\in \c}\left[ \frac{|E(C)|}{m}-\left(\frac{\sum_{v\in C} deg(v)}{2m} \right)^2 \right].
\end{equation}
where $E(C)$ is the set of edges connecting two vertices of the community $C\in \c$.


\bigskip

Given a graph $G$, the network coloring problem is that of coloring the vertices of a network so that no two adjacent vertices share the same color.  The smallest number of colors needed to color a graph $G$ is called its chromatic number, $\x(G)$. Graph coloring is computationally hard. Especially, it is NP-hard to compute the chromatic number. However, there are several algorithms, even parallelizable \cite{BE09}, which allow to color a graph with a number of colors upper bounded by $deg(G)+1$.

\section{Related Work} \label{sec:RW}
\subsection{Community Detection}
The goal of community detection algorithms is to partition a given network into communities consisting of nodes with similar characteristics. Specifically, a community is generally defined as a subset of nodes densely interconnected relatively to the rest of the network.
Several algorithm have been proposed which are typically classified in:  \textit{divisive} \cite{GN02}, \textit{agglomerative}\cite{NG04}  (depending on whether they focus
on the addition or removal of edges to or from the network) and \textit{optimization} \cite{BDGGHNW07} which continuously update the network partition in order to maximize a given measure of the quality of the network partition (i.e., the modularity). See \cite{DDDA05} and \cite{OL09} for detailed reviews.

\subsection{Label propagation algorithms}
The community detection strategy based on a label propagation algorithm (LPA), introduced by Raghavan et al. \cite{RAK07},  in contrast with previously proposed approaches, identifies network partitions  by an ``epidemic'' approach, i.e., it uses the diffusion of information in the network to identify communities.

\paragraph*{Main idea}
Initially, each vertex in the network is assigned a unique label, which will be used to determine the community it belongs to. Subsequently, an iterative process is performed
so that connected groups of vertices are  able to reach a consensus on some label giving rise to a community.
\\
At each step of the  process, each vertex updates its label to a new one which corresponds to the most frequent label among its neighbors. Formally, for each vertex $v \in V$,   $v$ updates it label according to 
\begin{equation}
 l_v= \argmax_{l} \sum_{u\in N(v)}[l_u==l]
\label{eq2}
\end{equation}
where  $l_v$ denotes the label of $v$ and
\begin{equation}
 \nonumber [P]=\begin{cases}1 & \hbox{ if  the statement } P \hbox{ is true}\cr		          
        					          0 & \hbox{ otherwise} 
					     \end{cases}
\end{equation}
denotes the Iverson bracket. When more than one choice is possible, ties are broken randomly (different ties management schemes  will be considered  in the rest  of the paper). This process is performed until some stop condition is met, e.g., no vertex changes its label  during one step. \\
A network community is then identified as a connected group of vertices having the same final label.

\medskip 

The process is formally described as Algorithm \ref{alg:sync}: Label Propagation (Synchronous). In the following, we denote by $l_v(i)$ the label of vertex $v$ at step $i$, for $i=0,1,\ldots$ and  $\forall \ v \in V.$ 
Initially all labels are distinct, that is $l_v(0)\neq l_w(0) \ \forall \ v, w \in V$. 
%
%
%

\begin{algorithm} [htb!]
\caption{\textbf{Label Propagation (Synchronous)} \label{alg:sync}}
\textbf{Initialize labels:} for each $v\in V, \  l_v(0)=v;$  \\
i=0;\\
\While{the stop criterion is not met}{
	i++;\\
	\textbf{Propagation:}\\
	\ForEach(){$v\in V$}{
 			$\displaystyle l_v(i)= \argmax_{l} \sum_{u\in N(v)}[l_u(i-1)==l],$
 	}
}
\textbf{return} Final labeling: $l_v(t)$ for each $v\in V$, where  $t$ is the last executed step.
\end{algorithm}

\subsection{Synchronous vs Asynchronous LPA} \label{synchvsasynch}
\begin{figure}[tb!]
\begin{center}
\includegraphics[width=0.40\textwidth]{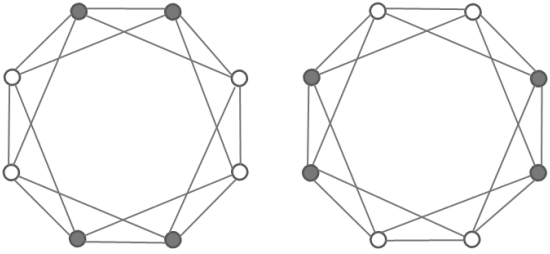}
\caption{The oscillation phenomenon on a non-bipartite network. Once one of the two configurations is entered, then the label values indefinitely oscillate between them.\label{es1}}
\end{center}
\end{figure}

The label propagation algorithm can be either synchronous, as presented above, or asynchronous. In the synchronous model (cf. Algorithm \ref{alg:sync}) each vertex computes its label at step $i$ based on the label of its neighbors at step $i-1$.
The synchronous algorithm is easy to implement and embarrassingly parallelizable. Since there are no dependencies between labels belonging to the same step, each \textit{Propagation} step can be executed in parallel over the vertices. 
However, it has been shown that synchronous updating may result in a cyclic oscillation of labels. This problem mainly occurs when the network considered contains a bipartite or a star-like component. We observe that label oscillations are not only due to bipartite or star subgraphs, several other kinds of graphs, see for instance Figure \ref{es1}, suffer this oscillation problem. 
In order to avoid possible cycles and ensure termination the authors of \cite{RAK07} suggest opting for an asynchronous approach (cf. Algorithm \ref{alg:async}).

\noindent
Although, the asynchronous approach deeply reduces the oscillation phenomenon, it  
exhibits some side effects: 
\begin{itemize}
\item
Since each vertex label is updated according to the current label of its neighbors, several dependencies need to be considered if one has to parallelize the algorithm: each vertex cannot compute his own label before each of its neighbor, which precedes it in the chosen permutation, has completed its computation. With more details, while the $foreach$ in line $6$ of Algorithm \ref{alg:sync} is parallelizable, this is not true for the  Algorithm \ref{alg:async}  (the computation on line $8$ depends on the current labels of the considered vertices). 
Parallelism is very important when the network size is large. Even if the LPA requires a few propagation steps, using the asynchronous approach, each step need to be sequenced among all vertices. 
Accordingly,  the amount of time needed by a parallelized version of the synchronous LPA scales logarithmically \cite{LHLC09}, whereas the time complexity of a parallel asynchronous LPA grows more than linearly, with respect to the network size.  
\item
Moreover, because during each iteration, the updating sequence is randomly chosen, the whole algorithm becomes unstable: different run of the algorithm may provide different final labeling.    
\item
A major limitation of the  asynchronous algorithm has been observed in \cite{LHLC09}: It often wrongly produces a ``monster'' community and several small communities. This problem is related to the fact that, due to the asynchronous nature of the algorithm, during the initial steps, the random permutation of the vertices tends to benefit the spread of some labels with respect to the others. For this reason certain communities do not form  links which are strong enough to prevent a foreign label flooding. Several experiments confirm that the synchronous version of the algorithm slows down the formation of such \textit{monster} communities, even though it does not completely prevent them \cite{LHLC09}.
\end{itemize}

\begin{algorithm}[tb!]
\SetCommentSty{footnotesize} 
\caption{ \textbf{Label Propagation (Asynchronous)} \label{alg:async}}
\textbf{Initialize labels:} for each $v\in V,  \ l_v(0)=v;$  \\
i=0;\\
\While{the stop criterion is not met}{
	i++;\\
	let $\pi=(v_{\pi(1)},v_{\pi(2)},\ldots,v_{\pi(n)})$ be a random permutation of the vertices.\\
	\textbf{Propagation:}\\
	\For{$j=1$ \emph{\KwTo} $n$}{
		$\displaystyle l_{v_{\pi(j)}}(i)= \argmax_{l} \sum_{u\in 	  N\left(v_{\pi(j)}\right)}[l_u==l]$, \\
		\hspace*{+0.3truecm}	where $l_u=\begin{cases} 
                                 l_u(i)  & \mbox{if }u=v_{ \pi(k)},\ k<j \\
   															l_u(i-1) & \mbox{if }u=v_{ \pi(k)},\ k>j
															\end{cases}$
  }
}
\textbf{return} Final labeling: $l_v(t)$ for each $v\in V$, where  $t$ is the last executed step.
\end{algorithm}

\subsection{LPA for bipartite networks}
A variation of the  LPA for bipartite networks was proposed in \cite{LM09}. The proposed algorithm has experimentally shown to be as effective as standard LPA, easily parallelizable, and stable (no label oscillations were observed).
Let $B$ (\textit{blue}) and $R$ (\textit{red}) be the two sets of vertices which correspond to the canonical partition of a bipartite network. The algorithm divides each propagation step into two synchronized stages. The first stage computes the new labels for blue vertices according to the current labeling of red vertices. When the labels of all the blue vertices have been updated, the second stage updates the label of red vertices according to the current labeling of blue vertices.

This algorithm corresponds to the asynchronous model where, for each propagation step, the permutation of vertices is such that blue vertices precede red ones.
Notice that, since the network is bipartite, both the stages are parallelizable: the label propagation for different red (resp. blue) vertices is independent of each other. 

\section{LPA Stopping Criteria and Tie Resolution Strategies}  \label{stopCriteria}
Before proceeding with the description of our proposal, we briefly discuss the stop criteria and tie resolution strategies for LPAs that will be considered in the following.
 
The simplest stop criterion   consists in  comparing the current labeling with the previous one, if  no label change occurs, the algorithm is stopped (successive step will not change any label). Unfortunately, the LPA, even in the asynchronous model, does not prevent cycles (i.e., a sequence of steps such that the final labeling corresponds to the initial one). In such cases the stop criterion defined above will fail. For this reason the stop criterion (c$1$) proposed in \cite{RAK07} is more elaborated: 

\medskip

\begin{minipage}{0.4\textwidth}  
\textit{if in the current labeling every vertex in the network has a label to which the maximum number of its neighbors belong to, then the algorithm is stopped.}
\end{minipage}
\hfill
\begin{minipage}{0.05\textwidth}
\begin{flushright}
 (c$1$)
\end{flushright}
\end{minipage}

\medskip

\noindent Said in other words, if in the next step all the labels change, if any, will come from ties, then the procedure ends and the current labeling determines a network partition. \\
Alternatively, in order to use a simple version of the stop criterion a change in the management of ties is required as proposed in \cite{BC09}. When a ties occurs (i.e., there are two or more labels that maximize the sum in the equation (\ref{eq2})) one of the labels that maximize the sum is chosen randomly. If, in case of ties, the current label of the vertex has priority over the others, the probability of generating cycles is sensibly reduced \cite{BC09}. We will call this version of the algorithm \textit{LPA with Precedence} (LPA-Prec).
Formally this variation is defined as follows: during the propagation step, if the current label satisfies  equation (\ref{eq2}), then the vertex keeps its current label, otherwise the algorithm follows the standard rules.
LPA-Prec is typically more stable then the LPA algorithm (a small amount of randomization is injected into the algorithm). However in some cases the LPA-Prec  stops whereas the standard LPA keeps running in search of better solutions. For this reason, in addition to be more stable, the LPA-Prec usually exhibits a faster convergence then LPA but the quality of the network partition generated can be a bit worse. 

Another tie resolution strategy guarantees the convergence of the algorithm. Assume that the set of labels allows defining a priority between each pair of labels. For instance, we can assume that labels are integer, and that a label $l$ has priority over a label $l'$ if $l > l'$. In this version of the  algorithm, named LPA-Max, each tie is solved deterministically by taking the label with higher priority between the set of labels that get the maximum in equation (\ref{eq2}).

Using results in \cite{PS83} one can show that 

\smallskip

\begin{fact} \label{fact} The synchronous version of LPA-Max does not generate cycle of size larger than two. 
\end{fact}

\smallskip

Fact \ref{fact}  implies  a simplification of the stop criterion (c$1$): 

\medskip

\begin{minipage}{0.35\textwidth}  
\textit{if either $l_v(i)=l_v(i-1)$ for each $v\in V$\\ or $l_v(i)=l_v(i-2)$ for each $v\in V,$\\ then the algorithm is stopped. 
}
\end{minipage}
\hfill
\begin{minipage}{0.05\textwidth}
\begin{flushright}
 (c$2$)
\end{flushright}
\end{minipage}

\medskip

Synchronous LPA-Max is completely deterministic; it will always generate the same network partition whenever  it starts with the same initial vertex labels. Anyway, by randomizing the initial labeling, or by using an asynchronous approach, it is possible to obtain different results.

Notice that when the number of initial labels is larger than $2$, the two variants of LPA can also be applied together, we will call this tie resolution strategy LPA-Prec-Max. 

\section{Semi--synchronous label propagation} \label{sec:ourResult}
In this Section we present the main contribution of this paper. Our proposal combines the advantages of both the synchronous and asynchronous model discussed above. Namely, our system is stable and efficient (easy parallelizable) as the synchronous model while does not undergo the oscillation problem.

We present a semi-synchronous LPA which allows to overcome the oscillation problem in  any network.  We will also formally prove  that our algorithm avoids  oscillations, that is, it converges to a stable labeling\footnote{Depending on the propagation rule used, some cycles can be still obtained due to the management of ties.}. 

Our work is inspired by the label propagation algorithm for bipartite networks given in \cite{LM09}.
We stress that, in general, the formal study of LPAs convergence is an open problem.
In particular,  no formal proof (or informal hint) that shows why the proposal in \cite{LM09} is able to defeat the oscillation issue has been provided.

We propose an  algorithm  which consists of two phases: 
\begin{itemize}
\item[1)]
{\bf Coloring Phase.} Color the network vertices so that no two adjacent vertices share the same color (i.e., by any distributed graph coloring algorithm). The coloring phase is easily parallelizable and efficient (only $O(deg(G))$ synchronous parallel steps) \cite{BE09};
\item[2)] {\bf Propagation Phase.} Each label propagation step  is divided into stages. 
Each stage is named upon  a different color. At  stage $c$, labels are simultaneously propagated to the vertices that have been assigned color $c$ during the coloring phase 1.
\end{itemize}

\begin{algorithm}
\SetCommentSty{footnotesize} 
\caption{ \textbf{LPA (Semi--synchronous) } \label{alg:our}}
\textbf{Initialize labels:} for each $v\in V,\  l_v(0)=v;$  \\
\textbf{Network coloring:} assign a color to the vertices of the network such that no two adjacent vertices share the same color. Let $\d=\{D_1,D_2,\ldots,D_\ell\}$ be the color partitioning obtained.\\
i=0;\\
\While{the stop criterion is not met}{
	i++;\\
	\textbf{Propagation:}\\
	\For{$j=1$ \emph{\KwTo} $\ell$}{
		\ForEach(){$v\in D_j$}{
 				$\displaystyle l_v(i)= \argmax_{l} \sum_{u\in N(v)}[l_u==l],$\\
			  \hspace*{+0.3truecm}	where $l_u=\begin{cases} 
                                 l_u(i)  & \mbox{if }u\in D_k,\ k<j \\
   															 l_u(i-1)  & \mbox{if }u\in D_k,\ k>j
															\end{cases}$
 		}
	}
}
\textbf{return} Final labeling: $l_v(t)$ for each $v\in V$, where  $t$ is the last executed step.
\end{algorithm}

A formal description of the algorithm is given as Algorithm \ref{alg:our}: LPA (Semi--synchronous).
It is easy to verify that the  number of stages per propagation step corresponds to the number of color needed to color the network, which is at most $deg(G)+1$. Moreover, each step of the propagation phase is easily parallelizable: since there are no dependencies between vertices having the same color, each stage can be executed synchronously. Hence the amount of time needed to reach a final consensus grows like  the number of propagation step times the number of stages per propagation step, where the former value has been experimentally shown to grow logarithmically with respect to $n$, while the latter is bounded by $deg(G)+1$.

\smallskip
\begin{theorem}
Consider a network $G=(V,E)$.
Assume that Algorithm  \ref{alg:our} uses the stop criterion (c$1$), that is, it ends at the first step $t$ such that  for each $v\in V$ one of the following condition holds
\begin{itemize}
	\item[i)]  $l_v(t)=l_v(t-1)$ 
	\item[ii)]  $l_v(t)\neq l_v(t-1)$ but this  change is due to a tie.
\end{itemize}
Then the Algorithm \ref{alg:our}  converges, independently of the tie management rule.
\end{theorem}
\smallskip

\begin{IEEEproof}{}
Let
$$
f(i)=\sum_{\{u,v\} \in E} [l_u(i)==l_v(i)]$$
be a function that computes the number of monochromatic edges of $G=(V,E)$ at step $i$, for any $i\geq 1$.
Clearly, the function $f(i)$ is upper bounded by $|E|$.
We will show that the value of $f(i)$ grows monotonically with the step number $i$.
Consider  any step $i$. If the stop criterion is not met, then at least one vertex $v$ has changed his label without the occurrence of a tie. Since the labeling occurs by colors and  neighboring vertices are assigned different colors, we get that  when a vertex $v$ is relabeled,  $v$'s neighbors do not change their labels. This immediately implies that  the number of monochromatic edges incident in $v$ strictly increases during step $i$. 
The same reasoning applies to each other vertex in $G$ that changes is label without the occurrence of a tie. Hence, we have that if after  label propagation step $i$ the stop criterion is not met, then $f(i)>f(i-1)$ and the result follows. 
\end{IEEEproof}

\begin{table*}[bt!] 
	\centering 
	\begin{tabular}{ccccc} 
		\hline\hline 
		$$ & \# of vertices \ \ & \# of Edges \ \ &  Modularity  (Known real partitioning) \ & Maximum modularity \\ 
		\hline\hline \\ 
		\textit{Karate} & $34$ & $78$ &  $0.383$ & $0.431$\\
		\hline \\[0.5ex]
		\textit{Dolphins} & $62$ & $159$ &  $0.492$ & $0.527$\\
		\hline \\[0.5ex]
		\textit{Football} & $115$ & $613$ &  $-$ & $0.588$\\
		\hline \\[0.5ex]
		\textit{NetScience} & $1,589$ & $2,742$ &  $0.955$ & $0.917$\\
		\hline \\[0.5ex]
		\textit{Power} & $4,941$ & $6,594$ &  $-$ & $0.807$\\
		\hline \\[0.5ex]
		\textit{Internet} & $22,963$ & $48,436$ &  $-$ & $0.516$\\
		\hline \\[0.5ex]
		\textit{Cond-Mat} & $31,163$ & $120,029$ &  $0.688$ & $0.657$\\
		\hline\hline 
	\end{tabular} 
 \caption{Networks' properties}
	\label{table:results} 
\end{table*}

\section{Experimental Results} \label{sec:Experiments}


We have implemented both the asynchronous and the semi--synchronous LPA with $4$ tie resolution strategies: LPA, LPA-Prec, LPA-Max and LPA-Prec-Max. We compared both the quality, the timing  and the stability of such algorithms on a set of real networks, for which we know the community structure.

\subsection{Analyzed Network and their properties}
The choice of networks to be used as a benchmark is a crucial problem. Several experiments have been executed on computer generated networks with a community structure known by construction \cite{GN02,PSSL09}. However generated networks cannot model real networks. Several studies have been developed in order to devise a class of benchmark graphs, having a known community structure and able to resemble real networks \cite{LFR08}. Results are rather preliminary and deserve much attention. Hence our tests have been performed over real networks having a known community structure \cite{internet}. Table \ref{table:results} reports some known properties of such networks:
\begin{enumerate}
	\item Zachary's Karate: Social network of friendships between $34$ members of a karate club at a US university in the 1970s \cite{karate};
	\item Dolphins: Social network of frequent associations between $62$ dolphins in a community living off Doubtful Sound, New Zealand \cite{dolphins};
	\item Football: Network of American football games between Division IA colleges during regular season Fall 2000 \cite{GN02};
	\item NetScience: Coauthorship network of scientists working on network theory and experiment, as compiled by M. Newman in May 2006 \cite{netscience};
	\item Power: A network representing the topology of the Western States Power Grid of the United State \cite{power};
	\item Internet: A symmetrized snapshot of the structure of the Internet at the level of autonomous systems, reconstructed from BGP tables posted by the University of Oregon Route Views \cite{internet};
	\item Cond-Mat: Coauthorships between scientists posting preprints between Jan 1, 1995 and June 30, 2003 on the Condensed Matter E-Print Archive \cite{condmat}.
\end{enumerate}

\subsection{Test Settings}
We have performed our tests by considering $56$ different test settings. Each test setting is characterized by the choice of the label propagation timing (asynchronous or semi--synchronous), the network and the tie resolution strategy.


Since LPAs involve a certain amount of randomization, we execute each test $100$ times and use the means and the variances of their ``performances'' for our comparisons. 

Each test setting is composed as follows:

\begin{itemize}
	\item[i)] We randomly assign the initial network labeling and  execute a greedy graph coloring algorithm which considers the vertices in increasing order of their labels. Notice that different initial assignment may correspond to different network coloring.
	\item[ii)] We execute LPA using the stop criterion (c$1$) proposed in \cite{RAK07}.  During each test we collect the number of phases required to reach the final labeling (\textit{timing}). 
	\item[iii)]  Then we use a simple propagation algorithm which identifies the communities as connected group of nodes having the same final label. Notice that disconnected group of nodes having the same label will be considered as different communities. 
	\item[iv)]  We  measure the average modularity of the partitioning obtained (\textit{quality}).  
	\item[v)] We also collect information about the standard deviation of the modularity of the partitioning obtained, the average number of communities and the size of the greatest community. Such data will be used to compare the stability of the algorithms (\textit{stability}). 
\end{itemize}

In the following each test setting is identified by a triple $(P,N,T)$ where
\begin{itemize}
	\item $P \in$ \{Asynchronous, Semi--synchronous\}, indicates the label propagation timing, 
	\item $N \in$ \{\textit{Karate, Dolphins, Football, NetScience, Power, Internet, Cond-Mat}\}, indicates the network and
	\item $T \in$ \{LPA, LPA-Prec, LPA-Max, LPA-Prec-Max\}, indicates the tie resolution strategy.
\end{itemize}

\subsection{Results}

\begin{figure*}
	\begin{center}
		\includegraphics[width=0.90\textwidth]{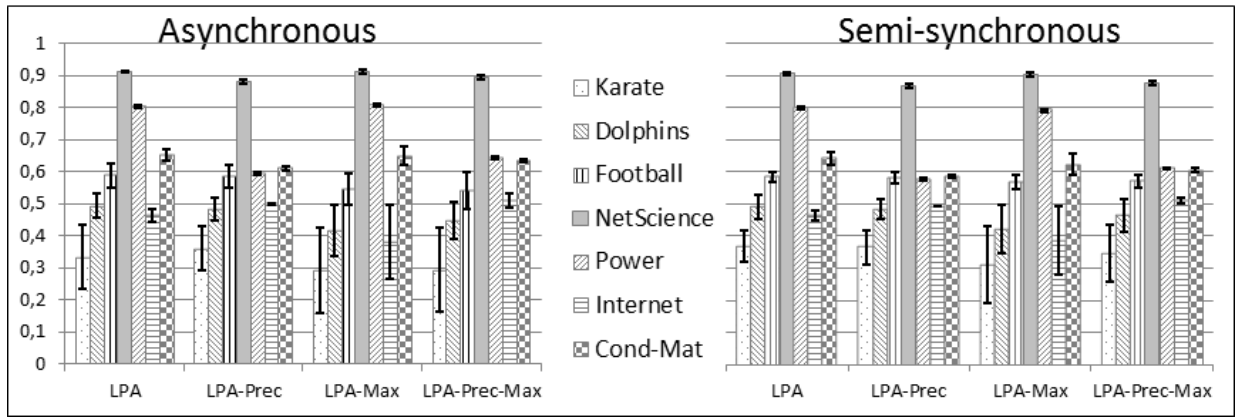}
		\caption{Average modularity and standard deviation measured on each test setting. Each test has been executed $100$ times.}
		\label{fig1}
	\end{center}
\end{figure*}

\noindent \textit{Quality:} Figure \ref{fig1} depicts the average modularity observed for each test setting. The results are quite similar, and reproduce the known values presented in Table \ref{table:results}. In particular the semi--synchronous approach slightly outperforms the asynchronous one on  \textit{Karate, Dolphins} and \textit{Football} networks while it is slightly worse on \textit{NetScience, Power} and \textit{Cond-Mat} networks. On the \textit{Internet} network performances are almost identical.  
We stress that our goal here is not to improve the quality of the results obtained by the asynchronous approach, rather we want to show that the semi-synchronous approach does not degrade any quality.

The accuracy of the results seems to be independent of the strategy used for the management of ties: except for the \textit{Power} network, where strategies LPA and LPA-Max provide a much better modularity ($\approx 0.8$) compared with LPA-Prec and LPA-Prec-Max ($\approx 0.6$), the other results are comparable.

\begin{figure*}
	\begin{center}
		\includegraphics[width=0.91\textwidth]{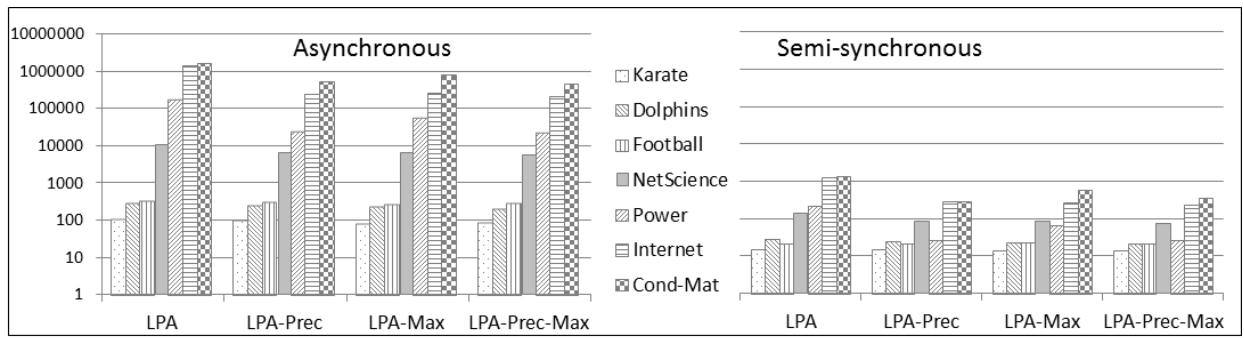}
		\caption{Average number of stages required by each test setting. The $y$-axis, which denotes the number of stages, is on a logarithmic scale.}
		\label{fig2}
	\end{center}
\end{figure*}

\medskip
\noindent \textit{Timing:} 
The efficiency of our proposal is shown in Figure \ref{fig2}. The improvement grows with the network size, in our tests it ranges from $\approx 5.7$: (Asynchronous, \textit{Karate}, LPA-Max) required $80$ stages, on average, while (Semi--synchronous, \textit{Karate}, LPA-Max) only $14$ stages, to $\approx 1802$:  (Asynchronous, \textit{Cond-Mat}, LPA-Prec) required $528772$ stages, on average, while (Semi--synchronous, \textit{Cond-Mat}, LPA-Prec) required $290$ stages. This means that our algorithm would greatly benefit from parallel implementations. Timing results also confirm that the number of steps needed to reach a final labeling grows logarithmically with respect to the size of the network and is independent from the model used (asynchronous or semi--synchronous). 

A careful comparison of this results  also shows  that all the variants of the standard LPA (that is, LPA-Max, LPA-Prec and LPA-Prec-Max) converge faster to the final labeling. As an example, the number of propagation steps required by (*, \textit{Power}, LPA) is $33.15$, on average, while (*, \textit{Power}, LPA-Max) is only $10.54$.  Notice that, in Figure \ref{fig2} results are shown with logarithmic scale in order to represent both the results of the asynchronous and semi-synchronous approaches. 

\medskip

\begin{figure*}
	\begin{center}
		\includegraphics[width=0.90\textwidth]{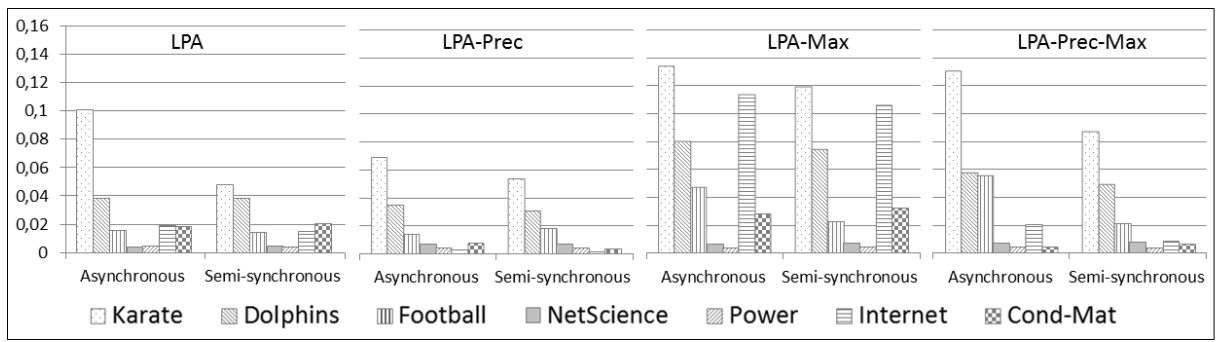}
		\caption{Standard deviation of the modularity measured on each test setting. Each test has been executed $100$ times.}
		\label{fig3}
	\end{center}
\end{figure*}

\noindent \textit{Stability:} In order to compare the stability of LPAs, we report, in Figures \ref{fig1} and  \ref{fig3}, the standard deviation ($\sigma$) of the modularity obtained for each test setting. The results indicate that the semi--synchronous approach is more stable than the asynchronous one.

Furthermore, by an accurate analysis of the results, it is possible to infer that whatever the algorithm used (semi--synchronous or asynchronous) the stability of results is strongly influenced by the strategy used for the management of ties. For this reason we show these results in four different graphs, one for each LPA variant (cf. Figure \ref{fig3}). 
In particular, the LPA-Prec is more stable than the others, while the LPA-Max is very unstable, especially in \textit{Karate, Dolphins, Internet} and \textit{Cond-mat} networks. The instability of LPA-Max is mainly due to the fact that before each test the initial labeling is randomized and consequently the way in which ties are solved changes too. On the other hand, LPA-Prec solves each tie in favor of its own label irrespective of the value of the labels involved.

\begin{figure}
	\begin{center}
		\includegraphics[height=3.6truecm]{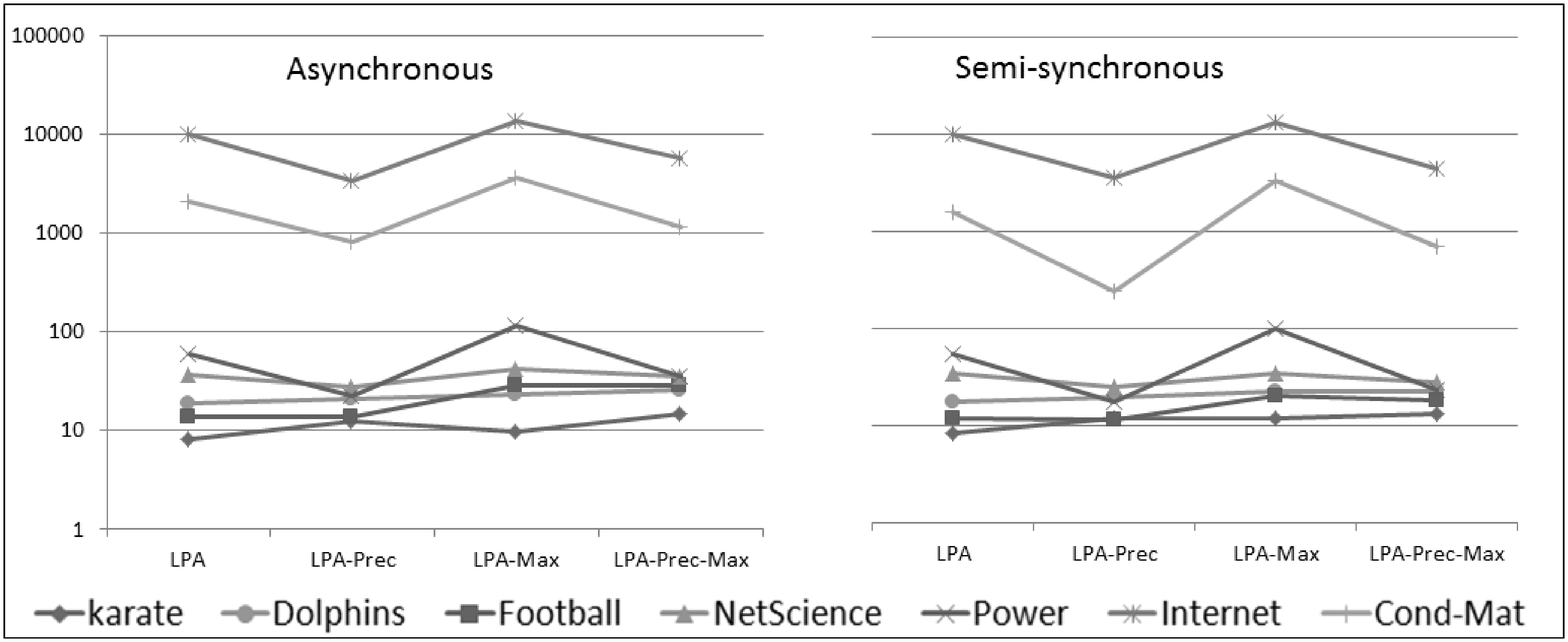}
		\caption{Average size of the biggest community.  The $y$-axis, which denotes the size of the biggest community, is on a logarithmic scale.}
		\label{fig4}
	\end{center}
\end{figure}

\begin{figure}
	\begin{center}
		\includegraphics[height=3.6truecm]{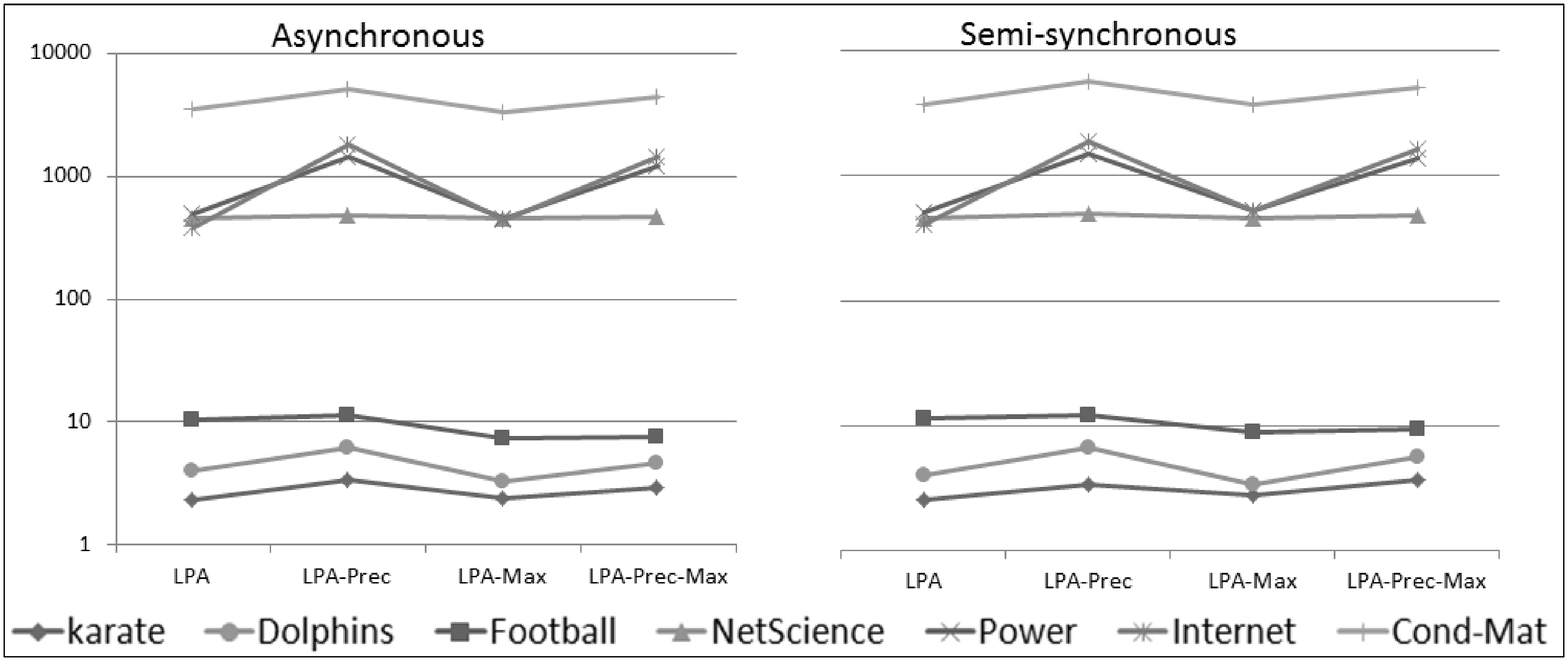}
		\caption{Average number of communities. The $y$-axis, which denotes the number of communities, is on a logarithmic scale.}
		\label{fig5}
	\end{center}
\end{figure}

We have also collected information about the size and the number of the communities generated during each test (cf. Figure \ref{fig4} and \ref{fig5}). The rationale behind this analysis is to understand whether the semi--synchronous algorithm avoids the generation of a ``monster'' community (cf. Section \ref{synchvsasynch}). The results confirm that the semi--synchronous LPA slightly reduces this phenomenon: the average size of the biggest community is always smaller while the average number of community is bigger. Results show also that several networks (e.g., \textit{Internet}) suffer this problem much more than others (e.g., \textit{NetScience}). 

To conclude we  also  note that even in this experiment, the stability of the algorithm is influenced by the strategy used to manage ties. All tests have shown that LPA-Prec is more stable than LPA and LPA-Prec-Max is more stable than LPA-Max.

\subsection{Discussion}
We have showed that our strategy provides good performances, with respect to several quality metrics. In particular we showed that our approach is easily parallelizable and would greatly benefit, in terms of convergence timing, of a parallel implementation. At the same time, neither the accuracy of the partitioning nor the stability of results are harmed by our strategy.

The set of developed tests also gives us the chance to perform a detailed comparison of $4$ different LPA tie resolution strategies. Results show that each strategy has some peculiarity: LPA-Max is faster than the other approaches, LPA-Prec is more stable and LPA-Prec-Max seems to be a good tradeoff. Results shows also that the peculiarity of LPA variants are not influenced by the approach (asynchronous or semi--synchronous) used during the label propagation step.

\section{Conclusion}
We have presented a novel semi--synchronized LPA for detecting network communities. Our proposal has shown to be 
\begin{itemize}
	\item \textbf{effective:} the accuracy of our approach is comparable with standard LPAs;
	\item \textbf{efficient:} it is easy to parallelize and consequently much faster than asynchronous approaches; 
	\item \textbf{stable:} our proposal has the advantage of ``limiting'' randomization to such an extent that results are quite uniform.
\end{itemize}

We showed that using a quite simple stop criterion, our algorithm is able to defeat the oscillation phenomena that preclude the use of synchronous strategies. 

\medskip

Several problems still remain open. In particular, characterizing the convergence of LPAs and determining  its speed is an open question. For instance, it has been shown that synchronous LPA-Max  converges to a period $2$ \cite{PS83}. What about the asynchronous and semi-synchronous approaches? What about the LPA-Prec and the LPA-Prec-Max variants?


\bibliographystyle{alpha}
\bibliography{lpbib}

\end{document}